\documentclass[a4paper,11pt]{article}
\usepackage{pos}
\usepackage{amsmath,bm}
\usepackage{graphicx}
\usepackage{microtype}
\usepackage{float}
\newcommand{\dd}{\mathrm{d}}
\newcommand{\ptg}{p_{\gamma\perp}}
\newcommand{\bl}{\bm{l}_{\perp}}
\newcommand{\bp}{\bm{p}_{\gamma\perp}}

\title{Connecting High-Twist Expansion and Color Glass Condensate Dynamics}

\author[a]{Yu Fu}
\author[b,c,d]{Zhong-Bo Kang}
\author[e,f,g]{Farid Salazar}
\author[h]{Xin-Nian Wang}
\author[i,j,k]{Hongxi Xing}

\affiliation[a]{Department of Physics, Duke University, Durham, North Carolina 27708, USA}

\affiliation[b]{Department of Physics and Astronomy, University of California, Los Angeles, CA 90095, USA}
\affiliation[c]{Mani L. Bhaumik Institute for Theoretical Physics, University of California, Los Angeles, CA 90095, USA}
\affiliation[d]{Center for Frontiers in Nuclear Science, Stony Brook University, Stony Brook, NY 11794, USA}

\affiliation[e]{Department of Physics, Temple University, Philadelphia, PA 19122-1801, USA}
\affiliation[f]{RIKEN-BNL Research Center, Brookhaven National Laboratory, Upton, New York 11973, USA}
\affiliation[g]{Physics Department, Brookhaven National Laboratory, Upton, New York 11973, USA}

\affiliation[h]{Key Laboratory of Quark and Lepton Physics (MOE) and Institute of Particle Physics, Central China Normal University, Wuhan 430079, China}

\affiliation[i]{State Key Laboratory of Nuclear Physics and Technology, Institute of Quantum Matter, South China Normal University, Guangzhou 510006, China}
\affiliation[j]{Guangdong Basic Research Center of Excellence for Structure and Fundamental Interactions of Matter, Guangdong Provincial Key Laboratory of Nuclear Science, Guangzhou 510006, China}
\affiliation[k]{Southern Center for Nuclear-Science Theory (SCNT), Institute of Modern Physics, Chinese Academy of Sciences, Huizhou 516000, China}

\abstract{
The high-twist (HT) expansion and the Color Glass Condensate (CGC) describe nuclear multiple scattering in complementary kinematic regimes. For direct-photon production in proton--nucleus collisions, we formulate an arbitrary-order multiple-scattering master expression that includes photon emission before, between, or after the ordered quark scatterings in the amplitude and its conjugate. In the high-energy limit, resumming multiple scatterings to all orders recovers the CGC Wilson-line dipole result. However, the naive power expansion of the eikonal CGC result reproduces leading twist but only part of the twist-four contribution obtained in the HT expansion. Including longitudinal sub-eikonal phases and Landau--Pomeranchuk--Migdal interference restores the CGC--HT correspondence through twist-4.
}

\begin{document}
\maketitle

\section{Introduction}

Multiple parton scattering in nuclei is essential for understanding nuclear-dependent phenomena in high-energy collisions. The high-twist (HT) expansion approach based on generalized QCD collinear factorization~\cite{Qiu:1990xxa} and the Color Glass Condensate (CGC) effective theory~\cite{McLerran:1993ni} describe such scatterings in different kinematic regions. In the HT approach, additional soft rescatterings are treated order by order as power corrections to leading twist, with nuclear properties encoded in multiparton correlation functions. At sufficiently small $x$, where the gluon occupation number becomes large, the CGC provides an effective description of the saturated regime. In this framework, nuclear properties are encoded in correlators of light-like Wilson lines, which resum multiple eikonal scatterings.

Although the two formalisms have different domains of validity, one may expect them to agree in the kinematic region where both are applicable. Direct-photon production in proton--nucleus collisions provides a simple but useful process for investigating this correspondence, since the photon does not undergo strong interactions with the nuclear medium. In this proceeding, we first present an arbitrary-order multiple-scattering expression and its coherent small-$x$ limit, and then show the mismatch between HT expansion and the naive power expansion of CGC. We finally exhibit the consistency between HT and CGC by including nuclear-enhanced initial- and final-state double scatterings and their interference, going beyond the shock-wave approximation and incorporating the Landau--Pomeranchuk--Migdal (LPM) interference effect.

\section{Arbitrary-Order Multiple Scattering and the CGC Limit}

\begin{figure}[h]
  \centering
  \includegraphics[width=0.9\textwidth]{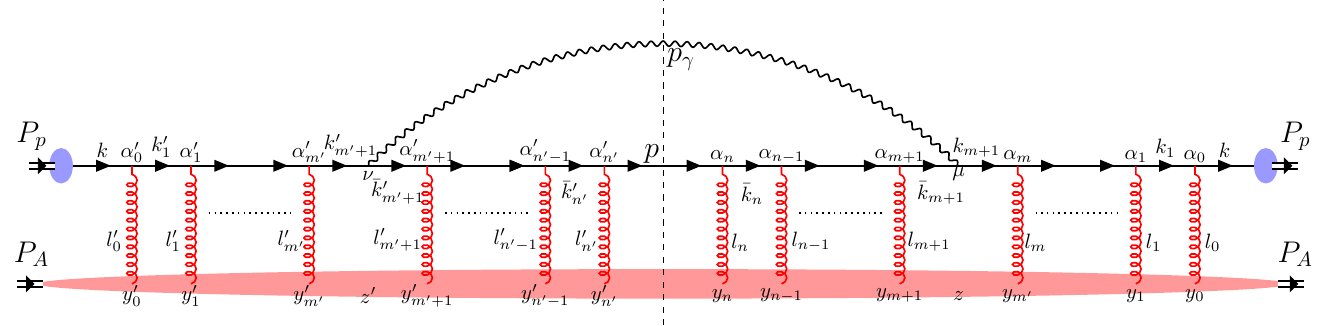}
  \caption{Representative  contribution with
  arbitrary $n+1$ and $n'+1$ ordered medium gluon insertions.}
  \label{fig:alltwist}
\end{figure}
We consider direct-photon production in $p(P_p)+A(P_A)\to\gamma(p_\gamma)+X$, focusing on the quark-initiated process $q(k)+A(P_A)\to q(p)+\gamma(p_\gamma)+X$. Here $P_A$ denotes the average momentum per nucleon, $k^-=x_qP_p^-$, and $\xi=p_\gamma^-/k^-$ is the longitudinal momentum fraction carried by the photon. The quark undergoes multiple scattering inside the medium, as illustrated in Fig.~\ref{fig:alltwist}. In general, there are $n+1$ medium gluon insertions with momentum transfer $l_i$ $(i=0,\ldots,n)$ in the amplitude. The label $m=-1$ places the photon before medium scattering, $0\leq m<n$ between insertions $m$ and $m+1$, and $m=n$ after medium scattering. Primed variables are analogous and belong to the conjugate amplitude. Following Refs.~\cite{Fries:2002mu,Majumder:2007hx}, we work in a covariant gauge and retain the leading $A^+$ component of the nuclear background, $A^\alpha\simeq A^+P_A^\alpha/P_A^+$. The contribution in Fig.~\ref{fig:alltwist} then reads
\begin{align}
E_\gamma\frac{\dd^3\sigma_{qA}^{nn';mm'}}{\dd^3\bm p_\gamma}
 &={\cal N}_{nn'}\int[\dd\Pi_{nn'}]\,
 \Phi_{nn'}^\perp{\cal C}_{nn'}{\cal H}_{nn'}^{mm'}
 I_{n,m}I'_{n',m'},
\label{eq:master-sum} 
\end{align}
where the factor ${\cal N}_{nn'}$ collects the flux, spin/color averages, couplings, and photon phase space, while ${\cal H}_{nn'}^{mm'}$ contains the quark-line trace, propagators, and the on-shell Jacobian for the final-state quark.

\begin{align}
 {\cal N}_{nn'}=\frac{g^{n+n'+2}e_q^2e_{\rm em}^2(P_A^+)^{n+n'+1}}
 {8N_c(2\pi)^3(k+P_A)^2}, \quad
 {\cal H}_{nn'}^{mm'}=
 \frac{2[1+(1-\xi)^2]\,\bm R_m\!\cdot\!\bm R'_{m'}}
 {(P_A^+)^{n+n'+3}k^-\xi^2(1-\xi)^2}.
 \label{eq:normalization-hard}
\end{align}
Here $e_{\rm em}$ is the elementary electric charge, and $e_q$ denotes the electric charge fraction of the quark. We define $K_i=\sum_{s=0}^{i}l_s$ ($K_{-1}=0$), and $\bm R_m=\bp-\xi\bm K_{m\perp}$, $\bm R'_{m'}=\bp-\xi\bm K'_{m'\perp}$, and $\bm K_{-1\perp}=\bm K'_{-1\perp}=0$.

By translational invariance, we set $y'_0=0$. Overall momentum conservation fixes $\bm l'_{n'\perp}$, leaving only $n'$ independent conjugate-side transverse transfers. The integration measures and the Fourier phase are
\begin{align}
 [\dd\Pi_{nn'}]&=
 \prod_{i=0}^{n}\!\left(\dd y_i^-\,\dd^2\bm y_{i\perp}
 \frac{\dd^2\bm l_{i\perp}}{(2\pi)^2}\right)
 \prod_{j=1}^{n'}\!\left(\dd y_j^{\prime-}\dd^2\bm y'_{j\perp}\right)
 \prod_{j=0}^{n'-1}\frac{\dd^2\bm l'_{j\perp}}{(2\pi)^2},
 \nonumber\\
 \Phi^\perp_{nn'}&=
 \exp\!\left[i\sum_{i=0}^{n}\bm l_{i\perp}\!\cdot\!\bm y_{i\perp}
 -i\sum_{j=1}^{n'}\bm l'_{j\perp}\!\cdot\!\bm y'_{j\perp}\right],
 \quad
 \bm l'_{n'\perp}=\bm K_{n\perp}
 -\sum_{j=0}^{n'-1}\bm l'_{j\perp}.
 \label{eq:measure-phase}
\end{align}
The medium correlation is 
${\cal C}_{nn'}=
\operatorname{Tr}_c\![
t^{a_n}\cdots t^{a_0}t^{a'_0}\cdots t^{a'_{n'}}]
\langle P_A|
\prod\limits_{i=0}^{n}A^{+,a_i}(y_i)
\prod\limits_{j=n'}^{0}A^{+,a'_j}(y'_j)
|P_A\rangle$
, with fundamental generators $t^a$.

We define the longitudinal pole positions
$\bar x_D^i=\bm K_{i\perp}^2/(2P_A^+k^-)$ and
$\bar x_C^i=\bp^2/(2P_A^+k^-\xi)
+(\bm K_{i\perp}-\bp)^2/[2P_A^+k^-(1-\xi)]$,
corresponding to quark propagators before and after
photon emission, respectively, together with
$x_L=\bp^2/[2P_A^+k^-\xi(1-\xi)]$.
With $x_i=l_i^+/P_A^+$ and $X_i=\sum_{s=0}^{i}x_s$, the pole factors are
$D_i^-=(X_i-\bar x_D^i-i\epsilon)^{-1}$,
$C_i^-=(X_i-\bar x_C^i-i\epsilon)^{-1}$,
and $C_{-1}^-=(-x_L-i\epsilon)^{-1}$.
The amplitude pole product is 
$\mathcal P_{n,m}^{-}
=
\left(\prod_{i=0}^{m}D_i^{-}\right)
\left(\prod_{i=m}^{n-1}C_i^{-}\right)$, with $-1\le m\le n$,
where empty products (i.e. $m=-1$) are understood as unity. The conjugate pole product ${\cal P}_{n',m'}^+$ uses primed variables and the opposite $+i\epsilon$ prescription. The longitudinal integrals are
\begin{align}
I_{n,m}
&=\int\prod_{i=0}^{n}\frac{\dd x_i}{2\pi}\,
(2\pi)\delta(X_n-\bar x_C^n)
e^{-iP_A^+\sum_i x_i y_i^-}
{\cal P}_{n,m}^-,
\nonumber\\
I'_{n',m'}
&=\int\prod_{j=0}^{n'}\frac{\dd x'_j}{2\pi}\,
(2\pi)\delta(X'_{n'}-\bar x_C^{\prime n'})
e^{+iP_A^+\sum_j x'_j y_j^{\prime-}}
{\cal P}_{n',m'}^+.
\label{eq:residues}
\end{align}

Starting from Eq.~\eqref{eq:master-sum} and summing over the photon emission positions $-1\le m\le n$ and $-1\le m'\le n'$ gives the contribution at fixed $(n,n')$. The full multiple-scattering result is then obtained by summing over $n,n'\geq 0$, where $n=0$ corresponds to one gluon insertion.

In the high-energy limit relevant for coherent small-$x$ scattering, the longitudinal phase differences between successive scatterings become negligible, $P_A^+(\bar x_C^m-\bar x_D^m)(y_{m+1}^--y_m^-)\ll 1$. The resulting LPM interference suppresses photon emission between scatterings relative to initial- and final-state radiation. Combining the four surviving endpoint contributions, $(m,m')=(-1,-1),\,(-1,n'),\,(n,-1),\,(n,n')$, including their interference, gives the photon radiation kernel ${\cal R}_\gamma(\xi,\bp,\bm K_\perp)
=
\frac{[1+(1-\xi)^2]\xi^2\bm K_\perp^2}
{\bp^2(\bp-\xi\bm K_\perp)^2}$.

The transverse momentum integrations set $\bm y_{i\perp}=\bm y_\perp$ and $\bm y'_{j\perp}=\bm y'_\perp$ for all gluon insertions in the amplitude and its conjugate, respectively. We use the correspondence between the normalized nuclear matrix element and the CGC average, $\langle{\cal O}\rangle = \frac{\langle P_A|{\cal O}|P_A\rangle}{\langle P_A|P_A\rangle} = \frac{\langle P_A|{\cal O}|P_A\rangle}{2P_A^+{\cal V}}$, where ${\cal O}$ denotes the ordered product of background fields and ${\cal V}$ is the normalization volume associated with translational invariance. The longitudinally ordered multiple scatterings then resum into the fundamental light-like Wilson line $V(\bm y_\perp) = {\cal P}\exp\left[ ig\int_{-\infty}^{\infty}\dd y^-\, A_a^+(y^-,\bm y_\perp)t^a \right]$. Since the series starts with one background-field insertion, the sum gives $V-\mathbf{1}$. For a translationally invariant target and a full transverse-plane Fourier transform, the identity terms have support only at zero transverse momentum transfer, where the photon radiation kernel vanishes for finite $\bp$. The cross section can thus be expressed in terms of the fundamental dipole correlator. Convolving with the proton quark distributions gives the standard CGC result~\cite{Gelis:2002ki,Fu:2024sba},
\begin{equation}
E_\gamma\frac{\dd^3\sigma_{pA\to\gamma X}}
{\dd^3\bm p_\gamma}
=
\frac{\alpha_{\rm em}}{2\pi^2}
\sum_q e_q^2
\int\dd x_q\,f_{q/p}(x_q)
\int\frac{\dd^2\bl}{(2\pi)^2}
F(x,\bl)\,
{\cal R}_\gamma(\xi,\bp,\bl).
\label{eq:cgc-photon}
\end{equation}
Here, $F(x,\bl)
=
\int\dd^2\bm y_\perp\,\dd^2\bm y'_\perp\,
e^{i\bl\cdot(\bm y_\perp-\bm y'_\perp)}
\left\langle
\frac{1}{N_c}
\operatorname{Tr}\left[
V(\bm y_\perp)V^\dagger(\bm y'_\perp)
\right]
\right\rangle_x$ is the Fourier transform of the fundamental dipole correlator, with $x$ the longitudinal momentum fraction probed in the nucleus.

Although the CGC result resums multiple eikonal scatterings to all orders, this resummation is not equivalent to a systematic high-twist expansion in inverse powers of the hard scale. To examine the connection between the two approaches, we compare the power expansion of the CGC result with the HT expansion through twist-4.

\section{Mismatch between HT and the power expansion of CGC}

In the HT formalism, nuclear multiple scatterings are organized as power corrections to leading twist, $\dd\sigma=\dd\sigma^{\rm LT}+\dd\sigma^{\rm T4}+\cdots$. For direct-photon production, the leading-twist and nuclear-enhanced twist-four contributions can be written schematically as~\cite{Guo:1995zk,Fu:2024sba}
\begin{align}
E_\gamma\frac{\dd^3\sigma^{\rm LT}}{\dd^3\bm p_\gamma}
&=f_{q/p}\otimes xf_{g/A}\otimes H_q^{(2)},\quad
E_\gamma\frac{\dd^3\sigma^{\rm T4}}{\dd^3\bm p_\gamma}
=f_{q/p}\otimes {\cal D}_X T_{gg}
\otimes H_{q+gg\to\gamma+q}^{(4)}.
\label{eq:ht-expansion}
\end{align}
Here $H_q^{(2)}=e_q^2\alpha_{\rm em}\alpha_s\xi^2[1+(1-\xi)^2]/(N_c\ptg^4)$, and $\otimes$ denotes convolution of different ingredients in the factorization.
The twist-four gluon-gluon correlation functions in the nucleus
can be written, for example, as
$T_{gg}(\{x_i\}) =
\int_{\{y_i^-\}}
e^{iP_A^+[x_1y^-+x_2(y_1^--y_2^-)+x_3y_2^-]}
\langle P_A|
F_\alpha^{\ +}(y_2^-)
F^{\beta+}(0)
F_\beta^{\ +}(y^-)
F^{+\alpha}(y_1^-)
|P_A\rangle$,
where $\int_{\{y_i^-\}}\equiv
(4\pi^2P_A^+)^{-1}\int dy^-\,dy_1^-\,dy_2^-
\theta(y^--y_1^-)\theta(-y_2^-)$.
For each cut diagram $X$, $\mathcal D_X$ denotes a polynomial
of up to second order in $\partial/\partial x_i$ acting on
$T_{gg}(x_1,x_2,x_3)$, including non-derivative contributions.

To compare with the HT result, we perform a power expansion in $1/\ptg^2$ of the CGC cross section in Eq.~\eqref{eq:cgc-photon} for $\ptg\gg Q_s$, where $Q_s$ is the nuclear saturation scale. Using the small-$x$ relations between the $\bl^2$ moment of the dipole distribution and the gluon PDF~\cite{Baier:2004tj}, and between its $\bl^4$ moment and the twist-four gluon correlation~\cite{Fu:2024sba}, we obtain
\begin{equation}
E_\gamma\frac{\dd^3\sigma^{\rm CGC}_{pA\to\gamma X}}
{\dd^3\bm p_\gamma}
=\sum_q\int\dd x_q\,f_{q/p}(x_q)H_q^{(2)}
\left[
xf_{g/A}(x)
+\frac{(2\pi)^2\alpha_s}{N_c}
\frac{4\xi^2}{\ptg^2}
T_{gg}(x,0,0)+\cdots
\right]_{x\to0}.
\label{eq:cgc-expansion}
\end{equation}
The first term reproduces the leading-twist result. The second term reproduces the non-derivative twist-four contribution if the longitudinal phases are neglected and the twist-four correlation functions reduce to the same small-$x$ limit. However, the complete HT result also contains derivative terms associated with longitudinal momentum transfers, which are absent in the eikonal CGC expansion. These terms account for the mismatch between the two formalisms at twist-4. For details of the calculation, see Ref.~\cite{Fu:2024sba}.

\section{CGC--High-Twist Correspondence at twist-4}

The mismatch in Section~3 arises from two ingredients neglected in the eikonal shock-wave approximation: longitudinal sub-eikonal phases and LPM interference. Retaining finite longitudinal momentum transfers and their associated phases, and expanding the Wilson lines in powers of $A^+$, allows a comparison with the HT formalism at fixed twist. For single scattering, the leading transverse derivatives convert two $A^+$ fields into $F^{+\perp}F_{\perp}^{\ +}$, reproducing the leading-twist gluon distribution, including its longitudinal phase~\cite{Fu:2023jqv}.

\begin{figure}[H]
 \centering
 \includegraphics[width=0.6\textwidth]{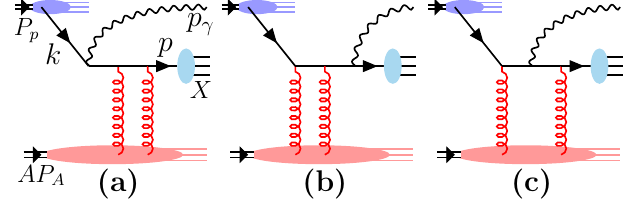}
 \caption{Initial- and final-state photon emissions
 (a,b) and intermediate emission (c).}
 \label{fig:lpm}
\end{figure}

At twist-4, photon emission between successive scatterings must also be included. As illustrated in Fig.~\ref{fig:lpm}, the eikonal approximation retains the initial- and final-state emissions (a,b), but neglects the intermediate emission (c). The two contributions to diagram (c) interfere through the phase difference~\cite{Fu:2023jqv}
\begin{equation}
1-\exp\left[i\frac{\Delta y^-}
{\tau_{\gamma,\mathrm{form}}}\right],
\qquad
\tau_{\gamma,\mathrm{form}}
=\frac{2k^-\xi(1-\xi)}
{(\bp-\xi\bm\ell_\perp)^2},
\label{eq:lpm}
\end{equation}
where $\Delta y^-$ is the separation between successive scatterings and $\bm\ell_\perp$ is the transverse momentum transfer before photon emission. When $\tau_{\gamma,\mathrm{form}}\gg\Delta y^-$, destructive LPM interference suppresses intermediate radiation. At finite $x$, however, the cancellation is incomplete, leaving a nonzero twist-four contribution.

To obtain the complete twist-four result, one must include both double scattering and single--triple interference contributions. In the transverse-momentum expansion, an additional pair of transverse derivatives converts the remaining $A^+$ fields into field strengths, yielding the twist-4 gluon correlation $T_{gg}$. The terms without these additional derivatives contribute to the leading-twist gauge link. Including the longitudinal phases and all relevant cut diagrams reproduces the derivative operators ${\cal D}_X$ of the HT calculation, together with the non-derivative contributions. The resulting correspondence schematically reads
\begin{equation}
\left.
E_\gamma\frac{\dd^3\sigma^{\mathrm{CGC}_{\rm sub}}}
{\dd^3\bm p_\gamma}
\right|_{\ptg>Q_s}
=
E_\gamma\frac{\dd^3(\sigma^{\rm LT}+\sigma^{\rm T4})}
{\dd^3\bm p_\gamma}
+\mathcal O(\ptg^{-8}).
\label{eq:match}
\end{equation}
Here $\sigma^{\mathrm{CGC}_{\rm sub}}$ denotes the background-field result including longitudinal sub-eikonal phases and LPM-sensitive contributions. For details of the calculation, see Ref.~\cite{Fu:2024sba}. Thus, retaining the sub-eikonal longitudinal phases and LPM interference establishes the CGC--HT correspondence through twist-4.

\section{Summary}

We have formulated an arbitrary-order multiple-scattering expression for direct-photon production in $pA$ collisions, including photon emission before, between, and after the scatterings. In the high-energy limit relevant for coherent small-$x$ scattering, summing over all scattering orders resums the multiple scatterings into light-like Wilson lines and recovers the CGC dipole result. The naive power expansion of this result reproduces leading twist but only part of the twist-four HT contribution. Including sub-eikonal longitudinal phases and LPM interference restores the CGC--HT correspondence through twist-4. Extending this correspondence to higher twists and other processes in $pA$ and $eA$ collisions offers a path toward a unified description of nuclear multiple scattering from dilute to dense regimes.

\section*{Acknowledgments}

This work was supported by the National Natural Science Foundation of China (NSFC) under Grants No.~12035007 and No.~12475139 (H.~X.) and No.~11935007 (X.-N.~W.); by the U.S. Department of Energy (DOE) under Grants No.~DE-FG02-05ER41367 (Y.~F.) and No.~DE-AC02-05CH11231 (F.~S., X.-N.~W.), and within the framework of the SURGE Collaboration (Z.-B.~K., X.-N.~W.); and by the U.S. National Science Foundation (NSF) under Grants No.~PHY-1945471 (Z.-B.~K., F.~S.) and No.~OAC-2004571 within the X-SCAPE Collaboration (F.~S., X.-N.~W.). F.~S. also acknowledges support from the Institute for Nuclear Theory under U.S. DOE Grant No.~DE-FG02-00ER41132.

\end{document}